\documentclass{article} 
\usepackage{emulateapj}
\usepackage[]{epsfig}
\hoffset -1.5truecm \lefthead{Menci et al.} 
\righthead{QSO Evolution in Hierarchical Structures} 
\def\newline{\hfil\break}
\begin{document}
\title{QUASAR EVOLUTION DRIVEN BY GALAXY ENCOUNTERS IN HIERARCHICAL STRUCTURES} 

\author{N. MENCI}
\affil{INAF - Osservatorio Astronomico di Roma,
via di Frascati 33, 00040 Monteporzio, Italy}

\author{A. CAVALIERE}
\affil{Dipartimento Fisica, II Universit\`a di Roma, via Ricerca Scientifica 1, 00133 Roma, Italy}

\author{A. FONTANA, E. GIALLONGO, F. POLI} 
\affil{INAF Osservatorio Astronomico di Roma,  via di Frascati 33, 00040 Monteporzio, Italy}

\author{V. VITTORINI}
\affil{Dipartimento Fisica, II Universit\`a di Roma,
 via Ricerca Scientifica 1, 00133 Roma, Italy} 

\begin{abstract}
We link the evolution of the galaxies 
in the hierarchical clustering scenario with the changing accretion rates of 
cold gas onto the central massive black holes that power the quasars. We 
base on galaxy interactions as main triggers of accretion; the related scaling 
laws are taken up from Cavaliere \& Vittorini (2000), and grafted to a 
semi-analytic code for galaxy formation. As a result, at high $z$  the 
protogalaxies grow rapidly by hierarchical merging; meanwhile, much fresh gas is 
imported and also destabilized, so the holes are fueled at their full Eddington 
rates. At lower $z$ the galactic dynamical events are mostly encounters in 
hierarchically growing groups; now the refueling peters out, as the residual gas 
is exhausted while the destabilizing encounters dwindle. So, with no parameter 
tuning other than needed for stellar observables, our model uniquely produces 
at $z>3$ a rise, and at 
$z\lesssim 2.5$ a decline of the bright quasar population as steep as observed. 
In addition, our results closely fit the observed luminosity functions of 
quasars, their space density at different magnitudes from $z\approx 5$ to 
$z\approx 0$, and the local $m_{BH}-\sigma$ relation.
\end{abstract}


\section {Introduction}

Observations pinpoint massive dark objects, conceivably supermassive 
black holes (BHs) with masses $m_{BH}\sim 10^6 - 5\,10^{9}$ $M_{\odot}$, at 
the center of most nearby bright galaxies (see Richstone et al. 1998). 
On the other hand, the space density of 
bright optical quasars (QSOs) is found to be lower by some $10^{-2}$ 
relative to the bright galaxies (see, e.g., Boyle et al. 2000). 

These observations support the view of the QSOs as a short active phase 
(lasting $\Delta t\sim 10^8$ yrs) of supermassive BHs which accrete 
surrounding gas at rates $\dot m_{acc}\sim 1-10^2$ $M_{\odot}$/yr,  see  
Rees (1984). If the accretion history starts from small primordial seeds (Madau 
\& Rees 2000), the accretion rate sets not only the QSO 
bolometric luminosities $L = \eta\,c^2\,\dot m_{acc}\lesssim 10^{48}$ 
erg/s (with standard mass-to-energy conversion 
efficiency $\eta\approx 0.1$, see, e.g., Yu \& Tremaine 2002), but also the relic BH masses.  

Additional observations relate the history of $\dot m_{acc}(z)$, and so the 
QSO evolution, with the galaxy structure and formation. 
These include the very magnitude of $\dot m_{acc}$ that needs to be drawn from 
the whole bulge of the hosts, and the correlation of $m_{BH}$ with 
the bulge luminosities (see Richstone et al. 1998), or better  with the  
velocity dispersion $\sigma$ (Ferrarese \& Merritt 2000; Gebhardt et al. 
2000). 

Relations of the sort have been explored in the 
framework of the hierarchical clustering scenario; here the galaxies condense from 
gravitationally unstable regions in the primordial  
density field of the dark matter (DM), then are assembled into poor groups and eventually into rich clusters. 
Such attempts are based either on analytic approaches (see Cavaliere \& Szalay 1986; Efstathiou \& Rees 1988; Carlberg 1990; Haehnelt \& Rees 1993; Haiman \& Loeb 1998; 
Burkert \& Silk 2001; Hatzminaoglou et al. 2002), or on semi-analytic models (SAMs) 
of galaxy formation (see Kauffmann \& Haehnelt 2000; Monaco, Salucci, Danese 2000; 
Volonteri, Haardt \& Madau 2002). But the fraction $f\equiv \Delta m_{acc}/m_c$ 
actually accreted out of the galactic gas remained  as a 
phenomenological quantity, described only in terms of scaling recipes, either 
containing tunable parameters or adjusted from the outset to the form of 
the observed correlations. Yet, all such models hardly accounted for the observed 
dramatic drop of the QSO population between $z\approx 2.5$ and the present. 

On the other hand, a physical law for the accretion has been proposed by 
Cavaliere \& Vittorini (2000, CV00); they suggested the amount of the 
cool galactic baryons accreted onto BHs to be given by the fraction $f$ 
destabilized and sent toward the nucleus by the gravitational torques arising either in major 
merging events (at $z\gtrsim 3$) or in  
encounters of the host with other galaxies inside a common DM halo (at $z\lesssim 2.5$). 
They use two analytic evaluations for the two regimes; in the latter 
they derive 
declining encounter rates and fast gas exhaustion concurring to produce 
an interestingly steep evolution of bright QSOs. 

Here we self-consistently compute halo and galactic quantities using a 
hierachical SAM of the kind proposed by Kauffmann et al. (1993);  
Somerville \& Primack (1999); Cole et al. (2000). Specifically, we insert the CV00 
model in the SAM developed by Menci et al. (2002); this includes galaxy interactions,  
and accounts for several statistical properties of the galaxy population like counts, 
$z$-resolved luminosity functions (LFs), sizes. This will allow us to 
continuously link the accretion 
history $\dot m_{acc}(z)$ and the QSO evolution with a fiducial model of galaxy 
formation consistent with the stellar observables. 

\section{Modeling the Galaxy Evolution}

The independent variables of our SAM (Menci et al. 2002) are: 
the circular velocity $V$ of the host DM halos (groups and 
clusters of galaxies with mass $M$ and virial radius 
$R$) containing the galaxies; the circular 
velocity $v$ of the DM clumps associated with the individual member galaxies. 
The former grow hierarchically to larger sizes through repeated merging events 
(at the rate given in Lacey \& Cole 1996), while the latter may coalesce 
either with the central galaxy in the common halo due to 
dynamical friction, or with other satellite galaxies through binary aggregations. 
The timescale for such galactic processes generally exceed the 
timescale for the merging of the host halos, so member galaxies accumulate in growing  
host halos.
  
At the cosmic time $t$ the SAM yields the following dynamical quantities.  
\newline
$\bullet$ The distribution $N(v,V,t)$ (per Mpc$^3$) of galaxies with circular velocity in 
the range $v - v+dv$ in host halos with circular velocity in the 
range $V - V+dV$. This is derived by computing iteratively the probability for a host halo 
to be formed from its progenitors, together with the 
probability that the member galaxies coalesce 
due to either dynamical friction or binary aggregations. As 
initial condition we assume a Press \& Schechter 
distribution, and we assign one galaxy to each host 
halo. From  $N(v,V,t)$ we derive the number $N_T(V,t)$ 
of galaxies in a host halo (membership), and the overall distribution 
of galaxy circular velocity $N(v,t)$ irrespective of the host halo. 
\newline
$\bullet$ The tidal radius $r_t(v)$, and the disk radius $r_d(v)$ and 
rotation velocity $v_d(v)$ computed after Mo, Mao \& White (1998). 
\newline
$\bullet$ The average relative velocity $V_{rel}(V)$ of the galaxies in a common DM halo. 

These dynamical quantities are associated to the baryons contained in 
the galactic DM halos in the way standard for SAMs. Initially the 
baryons contained in a galactic halo are in the amount $m\,\Omega_b/\Omega$ 
(with $m\propto v^3$ the DM mass of the galaxies)
and at the virial temperature; the mass $m_c$ of cold baryons is 
that residing in regions interior to the ``cooling 
radius''. Stars are allowed to form with rate 
$ \dot m_* = (m_c/t_{dyn})\,(v/200\,{\rm km\,s^{-1}})^{-\alpha_*}$, with the 
disk dynamical time evaluated as $t_d = r_d/v_d$. 
Finally, a mass $\Delta m_h=\beta\,m_*$ is returned 
from the cool to the hot gas phase due to the energy fed back   
by canonical type II Supernovae associated to $m_*$. The feedback efficiency is taken to be 
$\beta= (v/v_h)^{\alpha_h}$. The values adopted for the parameters 
$\alpha_*=-1.5$, $\alpha_h=2$ and $v_h=150$ km/s  
fit both the local B-band galaxy luminosity function and the Tully-Fisher relation, 
as illustrated by Menci et al. (2002). 
At each merging event, the masses of the different baryonic phases are refueled by 
those in the merging partner; the further increments $\Delta m_c$, $\Delta m_*$, $\Delta m_h$ 
from cooling, star formation and feedback are recomputed iterating 
the procedure described above.  

All computations are made in a $\Lambda$-CDM cosmology with 
$\Omega_0=0.3$, $\Omega_{\lambda}=0.7$, a baryon fraction 
$\Omega_b=0.03$, and Hubble constant $h=0.7$ in units of 100 
km s$^{-1}$ Mpc$^{-1}$. 

\section{BH Accretion triggered by Galaxy Encounters}
Here we just recall the key points of CV00, and recast their main 
equations in a form suitable for implementation in our SAM. 

For a galaxy with given $v$, the accretion of a fraction $f$ of the cold 
gas onto the central BH is intermittently triggered by interactions occurring at 
a rate 
\begin{equation} 
\tau_r^{-1}=n_T(V)\,\Sigma (v,V)\,V_{rel}(V)~. 
\end{equation}
Here $n_T=N_T/(4\Pi R^3/3)$, and the cross section 
%
$\Sigma(v,V)\approx\pi \langle \,(r_t^2+r_t^{'2})\rangle$ 
%
is averaged over all partners with tidal radius $r'_t$ in 
the same halo $V$. To this effect, 
the sizes, the relative velocities of the galaxies and the 
distribution of galaxy circular velocities in a common halo 
are previously computed from the SAM sector described in \S 2. 

The fraction of cold gas accreted by the BH in each interaction event is 
computed in eq. A3 of CV00 in terms the  variation 
$\Delta j$ of the specific angular momentum $j\approx Gm/v_d$ of 
the gas, to read
\begin{equation}
f(v,V)\approx {1\over 8}\,
\Big|{\Delta j\over j}\Big|=
{1\over 8}\Big\langle {m'\over m}\,{r_d\over b}\,{v_d\over V}\Big\rangle\, .
\end{equation}
Here $b=max[r_d,{\overline{d}(V)}]$ is the impact parameter, evaluated  
as the maximum between the galaxy $r_d$ and the average distance 
${\overline{d}}(V)=R/N_T^{-1/3}(V)$ of the galaxies in the halo. 
Also, $m'$ is the mass of the  partner galaxy in the interaction,  
and the average runs over the probability of finding a galaxy with mass $m'$ 
in the same halo $V$ where the galaxy $m$ is located. 
The prefactor accounts for the probability (1/2) of inflow rather than outflow 
related to the sign of $\Delta j$, and for the fraction (1/4) of the inflow actually reaching the 
nucleus rather than kindling circumnuclear starbursts, see Mirabel \& Sanders (1996).

Note from eqs. (1) and (2) that both the 
interaction rate $\tau_r^{-1}$ and the accreted fraction $f$ decrease with 
time, since in the growing host halos the increasing membership $N_T(V)$ is offset by 
the increasing $R$, $V$ and $V_{rel}$. In a group the values 
for $f$ range from a few $10^{-1}$ to several $10^{-3}$. The above 
behaviour of the accreted fraction $f$ is illustrated in the top panel of fig. 1. 

The average gas accretion rate triggered by interactions at $z$ is given by (CV00 eq. 5)
\begin{equation}
\dot m_{acc}(v,z)=\Big\langle{f(v,V)\,\,m_c(v)\over \tau_r(v,V)}\Big\rangle ~, 
\end{equation}
where the average is over all host halos with circular velocity $V$.
The bolometric luminosity so produced by the QSO hosted in a given galaxy 
is then given by 
\begin{equation}
L(v,t)={\eta\,c^2\Delta m_{acc}\over \tau} ~. 
\end{equation}
Here $\tau \approx t_d \sim 5\,10^7\,(t/t_0)$ yrs is the duration of the accretion 
episode, i.e., the timescale for the QSO to shine;  
$\Delta m_{acc}$ is the gas accreted at the rate given by eq. (3).
We take $\eta= 0.1$, 
and we obtain the blue luminosity $L_B$ by applying a bolometric correction 
of 13 (Elvis et al. 1994). 

The mass of the BH hosted in a galaxy with given $v$ at time $t$ is updated after 
\begin{equation} 
m_{BH}(v,t)=(1-\eta)\int_0^t\,\dot m_{acc}(v,t')\,dt' 
\end{equation}
assuming in all galaxies small seeds BHs of mass $10^2\,M_{\odot}$ 
(Madau \& Rees 2001); our results are insensitive to the specific
value as long as it is smaller than $10^5\,M_{\odot}$. 

\begin{center}
\vspace{-0.2cm} 
\scalebox{0.45}[0.45]{\rotatebox{0}{\includegraphics{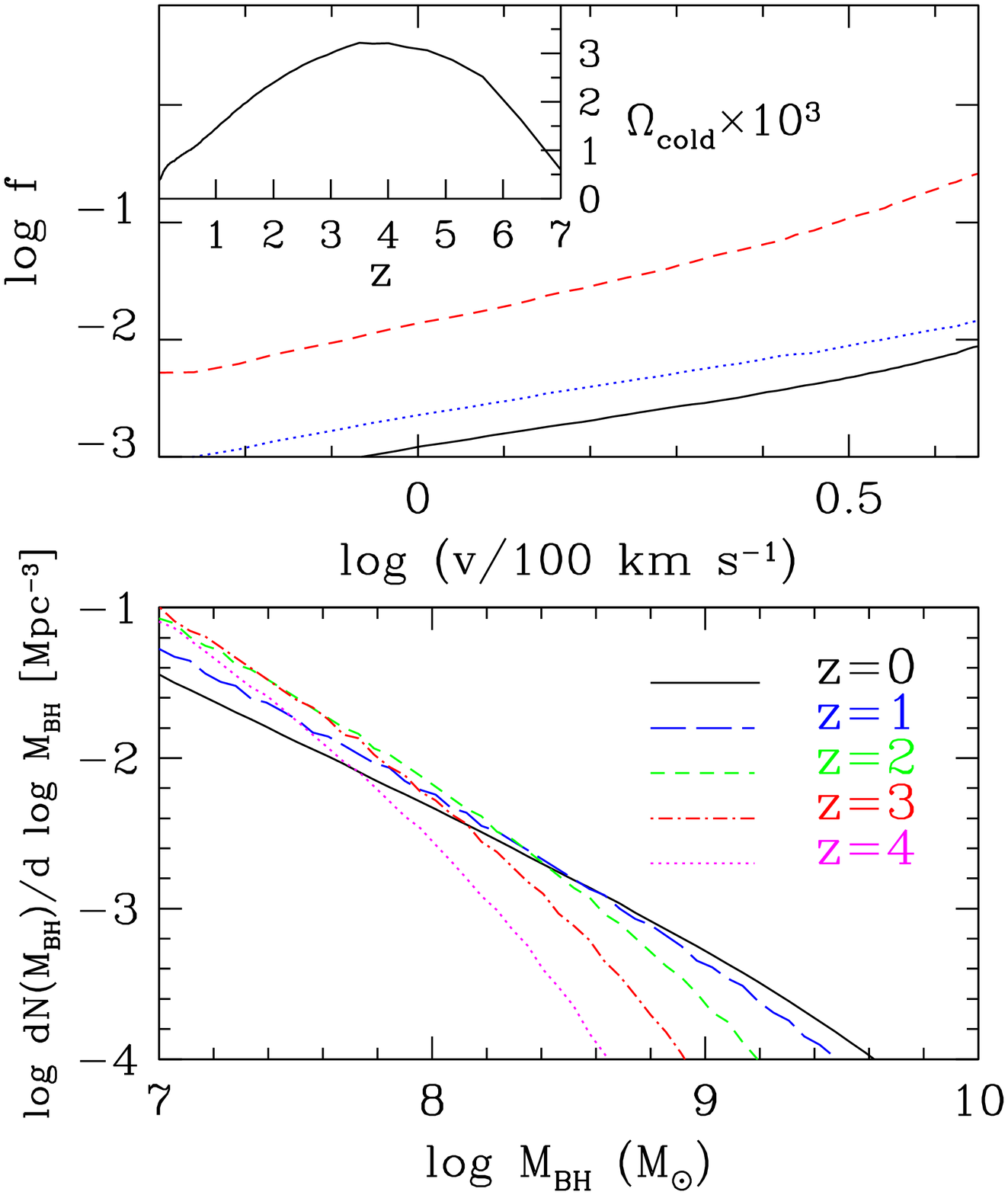}}}
\end{center}
{\footnotesize
\vspace{-1.5cm } 
Fig. 1. - Top: The fraction of cold gas accreted onto BHs as a function of 
the circular velocity of the host galaxy at $z=0.4$ (solid line), $z=1$ (dotted line) 
and $z=3$ (dashed line). The inset shows the 
the cosmological density of all the available cold gas in galaxies 
as a function of the redshift $z$.
Bottom: The mass function of BHs for various values of the redshift $z$.
\vspace{0.1cm}}
\section {Results}
In fig. 1 (bottom) we plot the mass function of BHs at 
four different redshifts. Note the mild evolution from $z=1$ 
to the present, due to the decrease of the accreted fraction 
$f$ (see top of fig. 1) and of the related $\Delta m_{acc}$ 
dut to the declining rate of merging and encounters (see \S 
3), and to the simultaneous exhaustion of the galactic cold 
gas $m_c(z)$ available for accretion (see inset in fig.1). 

The relation we obtain between the BH mass and the central 1-D 
velocity dispersion $\sigma$ of the host galaxy is shown in fig. 2. 
We use the canonical relation
$\sigma=v/\sqrt{2}$ approximately holding for an isothermal profile 
(see Binney \& Tremaine 1986). 

The steep slope of our relation (close to   
$M_{BH}\propto \sigma^4$ in the range $10^7\,M_{\odot}\lesssim M_{BH}\lesssim 
10^9\,M_{\odot}$) is the combined result of two processes: the merging histories of the 
galactic DM clumps, which by themselves would imply the mass of cold available gas  
to scale as $\sigma^{2.5}$; the destabilization of the cold gas by the 
interactions, which steepens the relation to $\sigma^{3.5}$. The further 
steepening to $\sigma^4$ is provided by the Supernovae feedback which 
depletes the residual gas content in shallow 
potential wells. No effort has been done to 
adjust the latter to match the relation to the data;
a slope steeper yet would be provided, especially in the upper range, 
by adding the feedback from the QSO emission itself onto the host gas 
(which does not alter the average galactic star formation in a group). 

\begin{center}
\vspace{-1.5cm} 
\scalebox{0.43}[0.43]{\rotatebox{0}{\includegraphics{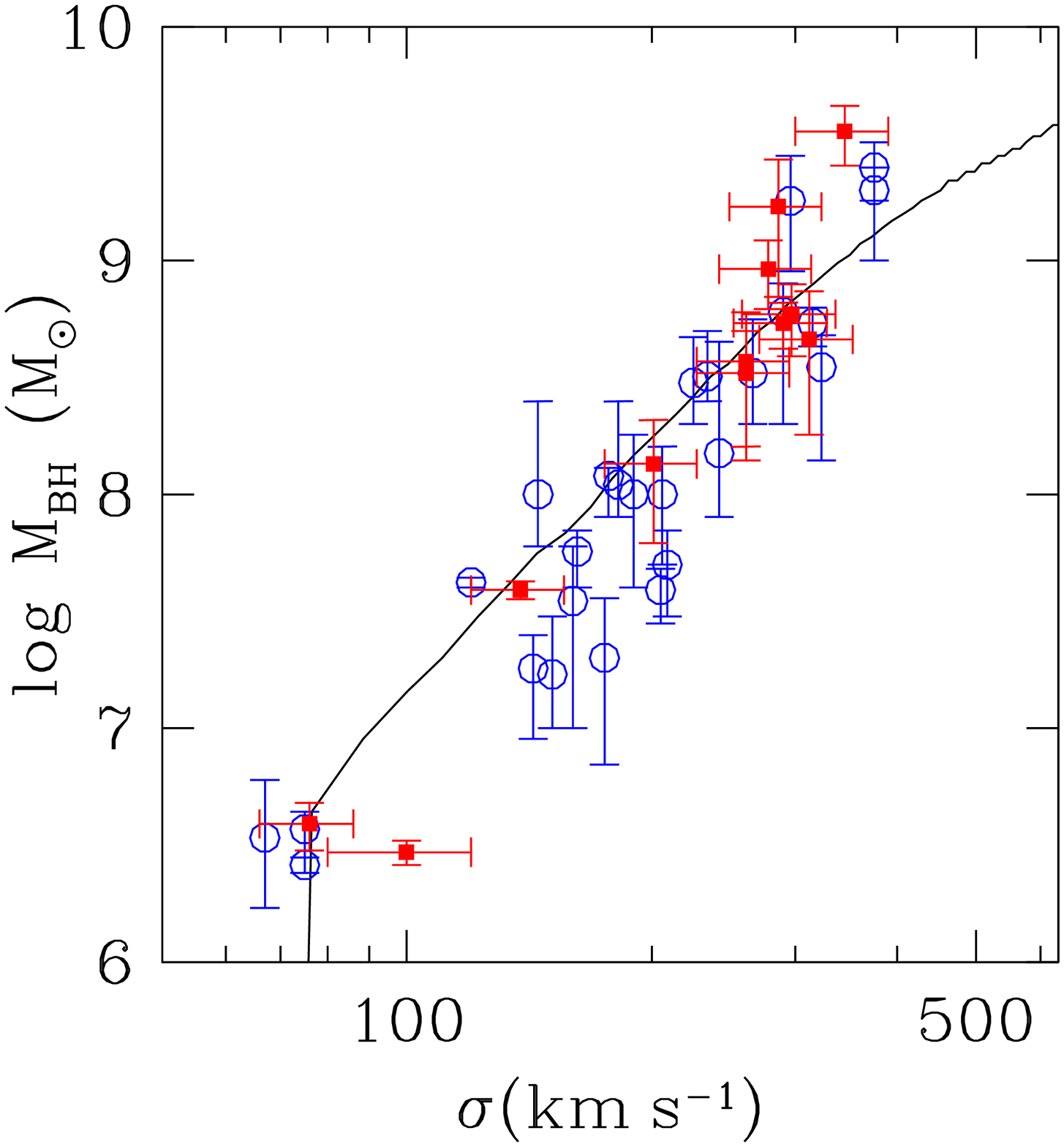}}}
\end{center}
{\footnotesize
\vspace{-1.3cm } 
Fig. 2. The derived relation between the BH mass and the 1-D velocity 
dispersion of the host galaxy (solid line) is compared with 
data from Ferrarese \& Merritt (2000, filled squares) and 
Gebhardt et al. (2000, circles).}
\vspace{0.3cm}

The evolving LF of the QSOs is derived from 
$N(v,t)$ by applying the appropriate Jacobian, see \S 3.  
The LF will include a factor $\tau /\langle\tau_r(v)\rangle 
< 1$ since the luminosities in eq. (5) last for a time $\tau$
and are rekindled after an average time $\tau_r$. The result is 
\begin{equation} 
N(L,t)=N(v,t)\,{\tau\over \langle\tau_r\rangle}\,\Big|{dv\over dL}\Big|
\end{equation} 
and is shown in fig. 3. 
Note the strong evolution from $z\approx 0.5$ to $z\approx 2$, due to both 
the declining $\Delta m_{acc}(z)$ and the timescale $\tau$ 
shortening at higher $z$.

In fig. 4 we plot  the predicted cosmic density $\rho_{Q}(z)$ of QSOs brighter than 
$M_B=-24$ and $M_B=-26$. The computed emissions, limited by the Eddington 
luminosity $L_E$, peak at $z \approx 2.5$; such redshifts are those where 
our SAM predicts the buildup of bright galaxy to peak (see Menci et al. 2002). 
Note that if no Eddinton limit were imposed, the 
density would be that represented by the dashed line in the bottom panel, which 
overshoots the data; this shows that at $z>2.5$ the QSOs do emit at their full 
$L_E$, due to the high merging rate which provides abundant refueling. 
At later times, the fueling is triggered mostly by the interactions with encounter rate 
declining as discussed in \S 3, and draws from scarce residual gas approaching exhaustion.  
This causes the dramatic fall of the QSO density we predict for  redshifts $z<2$. 
\begin{center}
\vspace{-1.3cm} 
\scalebox{0.43}[0.43]{\rotatebox{0}{\includegraphics{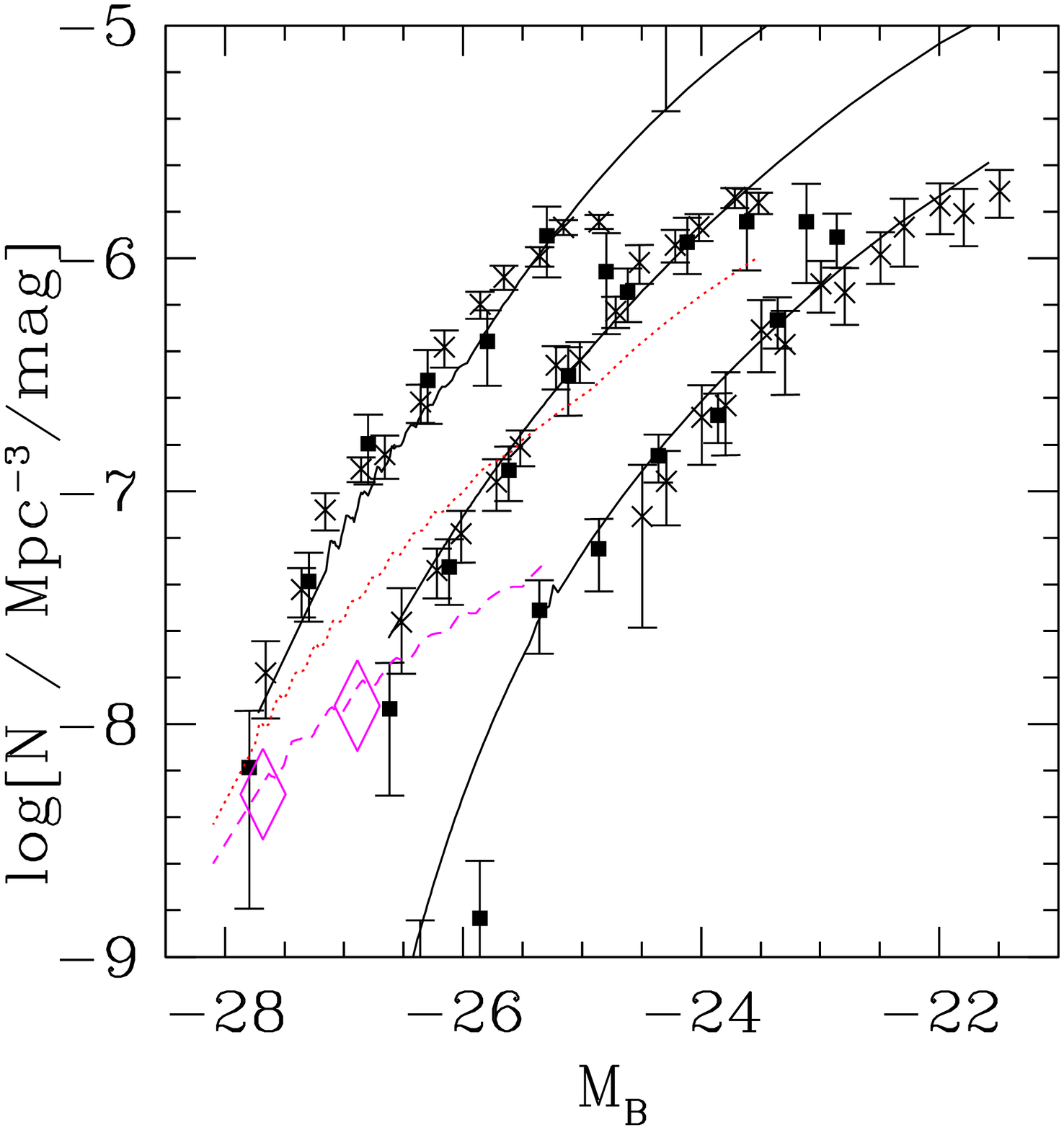}}}
\end{center}
{\footnotesize
\vspace{-1.4cm }
Fig. 3. The LFs from our model (solid lines) are shown for $z=0.55$ (lower curve), 
$z=1.2$ (middle curve) and $z=2.2$ (uppermost curve), and are compared with the 
data points. These are taken from Hartwick \& Shade (1990, solid squares) and 
Boyle et al. (2000, crosses), and rescaled to our cosmology with $\Omega_0=0.3$, 
$\Omega_{\lambda}=0.7$, $h=0.7$. We also show as a dashed line the model LF for 
$z=4.2$ compared to the Sloan data from Fan et al. (2001, diamonds). 
The dotted line is the predicted LF for $z=3.4$.}

\section{Conclusions} 

Our approach links the evolution of QSOs to that of their host galaxies on using 
scaling laws for the BH accretion based on triggering galaxy interactions, with  
no new parameters added to the galaxy formation model. 
It provides good fits to the observations concerning the 
$m_{BH}-\sigma$ correlation, the evolution of the QSO luminosity 
function, and their space density selected at different optical magnitudes. 
The agreement holds in the full range $z<6$ {\it continuously} covered by our model, and
supports the underlying picture that follows.  

At $z>2.5$ the QSOs emit at their full Eddington 
limit, because the high merging rate in this epoch of galaxy assemblage
insures both a high baryonic content in their hosts and an abundant BH fueling. 
As shown in fig. 3, in this range the LFs are {\it flat} (consistent with by recent data, see 
Fan et al. 2003) and {\it rise} with time. At $z<2.5$ 
the {\it steep drop} we find for the population  
results from three concurring processes: 
1) the declining rate of merging between early galaxies, which halts the 
acquisition of new gas available for accretion; 2) the progressive exhaustion of 
the baryon reservoirs in the hosts, consumed by  fast conversion 
into stars and by previous accretion episodes; 3) the 
eventual decline of the fraction $f\sim\Delta j/j$ of residual cold gas 
which is destabilized and accreted onto the central BHs by the dwindling   
interactions between galaxies. The latter (see fig.1) is at variance 
with the Kauffmann \& Haehnel (2000) model, and is crucial 
in matching the observed strong evolution of QSOs at $z < 2$; 
as a consequence, our model predicts an average Eddington ratio dropping 
from $L/L_E\sim 1$ at $z\approx 2.5$ to $L/L_E \sim 10^{-2}$ at $z\approx 0$, 
with a weak dependence on $m_{BH}$, consistent the with available data (see Woo \& Urry 2002). 

\vspace{-1.5cm}
\begin{center} 
\scalebox{0.43}[0.43]{\rotatebox{0}{\includegraphics{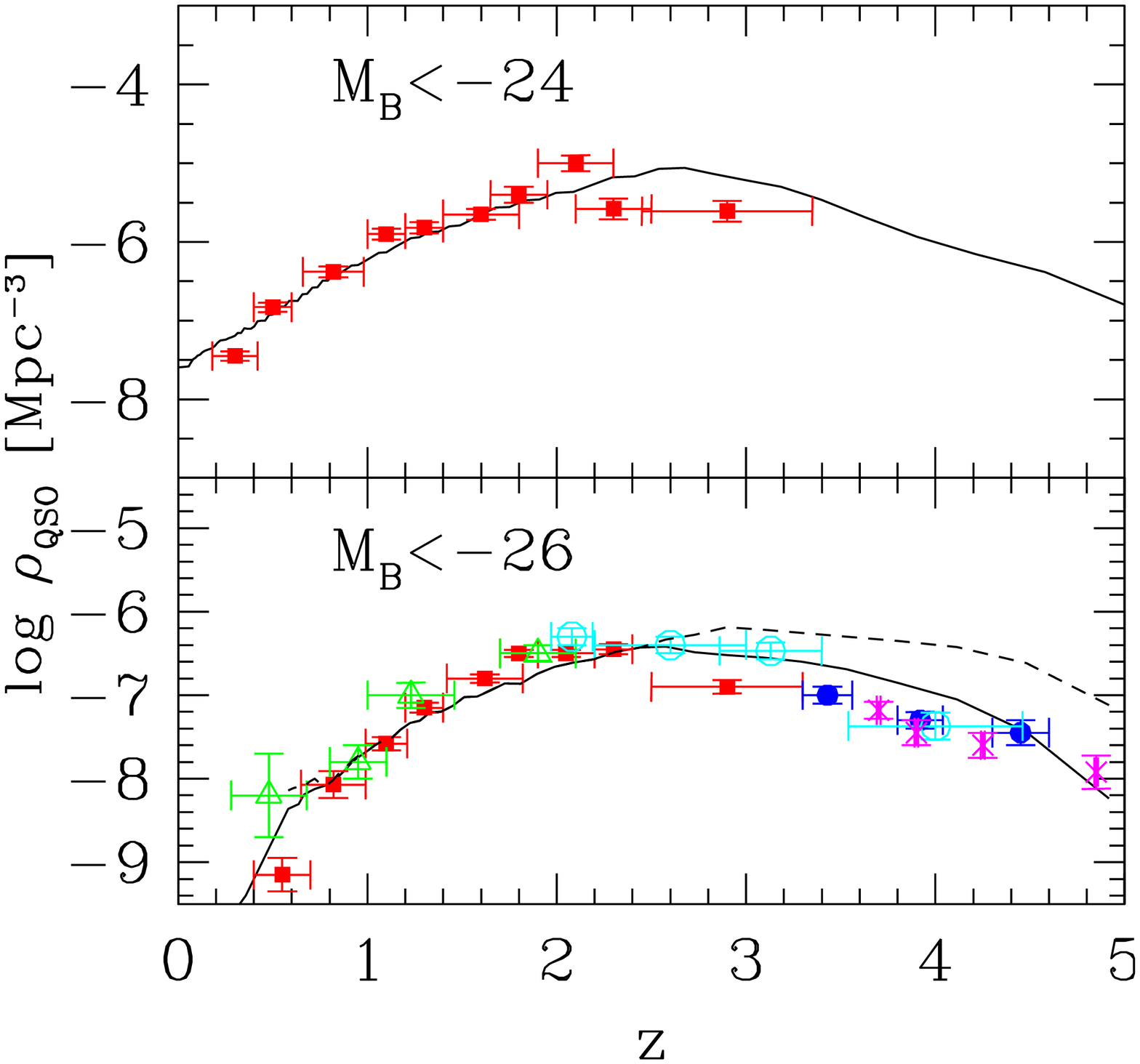}}}
\end{center}
{\footnotesize
\vspace{-1.8cm} Fig. 4. 
The predicted cosmic density of bright (upper) and very bright 
(bottom panel) QSOs. The dashed line represents the outcome when no Eddington 
limit were assumed (see text). The densities 
have been rescaled as appropriate for the critical cosmology with 
$h=0.5$ used by the authors for their (magnitude-limited) data; these are 
taken from Hartwick \& Shade (1990, solid squares); Goldschmidt \& Miller (1998, 
triangles); Warren, Hewett \& Osmer (1994, open circles); Schmidt, Schneider \& 
Gunn (1995, filled circles); Fan et al. (2001, crosses). }
\vspace{0.2cm}

The above picture implies a specific connection of the QSO 
emission with observable properties (magnitudes, colors and 
star content) of their host galaxies. On this we shall 
present elsewhere our results; we only stress that in our 
present model the host galaxies of bright QSOs have dynamic 
properties conducive to strong interactions, and hence to 
high luminosities, at early $z$. 


\end{document}